\documentclass[aps,prl,twocolumn,groupedaddress,showpacs,lengthcheck]{revtex4}

\usepackage{ams}
\usepackage[dvips]{epsfig}
\usepackage{subfigure}

\begin{document}

\title{Reflection, Transmission and Trapping Dynamics of Lattice Solitons at Interfaces}

\author{Y. Kominis}
\affiliation{School of Electrical and Computer Engineering, National Technical University of Athens,Zographou GR-15773, Greece}
\author{K. Hizanidis}
\affiliation{School of Electrical and Computer Engineering, National Technical University of Athens,Zographou GR-15773, Greece}

\begin{abstract}
Surface soliton formation and lattice soliton dynamics at an interface between two inhomogeneous periodic media are studied in terms of an effective particle approach. The global reflection, transmission and trapping characteristics are obtained in direct analogy to linear Snell's laws for homogeneous media. Interesting dynamics related to soliton power-dependent formation of potential barriers and wells suggest spatial filtering functionality of the respective structures. 
\end{abstract} 
\maketitle

The formation and the properties of spatially localized waves in periodic nonlinear structures, known as Lattice Solitons (LS), have been the subject of intense research interest, both theoretically and experimentally, in the areas of Nonlinear Optics \cite{JoViFa_07-Ru_03,PhysRep_08,DhLeSi_03-FlBaCo_05} as well as in Bose Einstein Condensates in optical lattices, \cite{BEC_1-2} while they are also closely related to solid state physics. In the context of Nonlinear Optics, these studies have led to the emergence of new research areas where novel phenomena are observed in photonic crystals, photonic crystal fibers, coupled waveguide arrays and optically induced lattices \cite{JoViFa_07-Ru_03,PhysRep_08,DhLeSi_03-FlBaCo_05}. Properties of wave localization and dynamics in such media with engineered characteristics suggest potential applications related to all-optical signal processing and optical circuits. \

The transverse inhomogeneity of these media results in breaking of the translational invariance. Thus, the existence of traveling waves is generally not ensured, while wave localization can take place in specific positions with respect to the underlying structure \cite{PeSuKi_04,SiFiIl_08}. Depending on the complexity of the structure the position and the stability of a LS may additionally depend on its power and width \cite{KoHi_08}. In addition to infinite periodic or quasiperiodic structures, LS can be formed at the interface between two semi-infinite media with different characteristics. It has been shown that such states can be formed at the boundaries between a periodic and a homogeneous medium (either linear or nonlinear) \cite{MaHuCh_06,KaVyTo_06,exp,KoPaHi_07}, as well as between two dissimilar periodic media \cite{KoPaHi_07,SuMaCh_0708}. These states, known as Surface Solitons (SS), are analogous of the so-called Tamm states in semiconductors.\
    
While the existence of stable and unstable stationary SS has been extensively studied in a variety of configurations, the investigation of LS and SS dynamics has not been explored yet. It is well known that, although the breaking of translational invariance does not allow for exact traveling wave solutions, LS can travel, without significant radiation emission, across an inhomogeneous structure under certain conditions for which the effective potential seen by a LS is weak. This can be either due to a small depth of the modulation of the respective property of the medium or due to the fact that the modulation period of the lattice is much shorter than the LS width.\

The present work is motivated by the interest on the reflection and transmission characteristics of such mobile LS at interfaces between different media and the relation of their dynamics with the existence of SS which are trapped in the corresponding boundaries. We study LS dynamics and SS formation at the interface between two dissimilar nonlinear periodic media (the case of an interface between a periodic and a homogeneous medium can be treated as a special case). A theory is developed, where the formation of SS as well as the reflection and transmission characteristics of a LS incident, at an angle, at the interface (separating two periodic media), is described analytically by an effective particle perturbation theory. The effective potential for each soliton determines the location and the stability type of each SS, which is different for waves of different power and width. More importantly, this potential determines the LS transmission characteristics in an analogy to Snell's law of geometric optics. Interesting nonlinear dynamics related to the presence of potential barriers and wells are demonstrated. The dynamical features are shown to depend strongly on the soliton characteristics, so that different solitons propagating at the same medium can possess qualitatively different dynamical behavior, suggesting a power-dependent spatial filtering mechanism.\

The underlying model describing wave propagation in an inhomogeneous, Kerr-type nonlinear medium is the perturbed Nonlinear Schr\"{o}dinger Equation (pNLSE)  
\begin{equation}
i\frac{\partial u}{\partial z}+\frac{\partial^2 u}{\partial x^2}+2|u|^2u-\epsilon n_0(x)u=0 \label{NLSE}
\end{equation}
where $z$ and $x$ are the normalized propagation distance and transverse coordinate, respectively, $\epsilon$ is a small dimensionless perturbation parameter, and $n_0(x)$ is the transversely inhomogeneous linear refractive index of the medium. The latter describes a structure consisting of two semi-infinite lattices interfaced at $x=0$ and having, in general, different amplitudes and periods 
\begin{equation}
n_0(x)=A(x)\sin\left[K(x)x+\phi\right] \label{n0}
\end{equation} 
with $\phi$ a constant phase, $A(x)=A^{(-)}+(\Delta A/2)[1+\tanh(ax)]$ and $K(x)=K^{(-)}+(\Delta K/2)[1+\tanh(ax)]$. The amplitude and the wavenumber of the left side lattice are $(A^{(-)},K^{(-)})$, while for the right side lattice we have $(A^{(+)},K^{(+)})=(A^{(-)}+\Delta A,K^{(-)}+\Delta K)$. The smoothness of the transition at the interface (at $x=0$) is determined by the parameter $a$. Note that for $A^{(-)}=0$ or $K^{(-)}=0$ we have the case of a homogeneous nonlinear medium interfaced with a nonlinear periodic lattice.\

For small perturbations (small $\epsilon$) the soliton is treated as an effective particle \cite{AcMoNe_89}, whose center of mass moves in an effective potential which determines the energy "landscape" seen by each soliton. In order to ensure robust soliton propagation close to the interface, we consider cases where the amplitude and wavenumber differences ($\Delta A$ and $\Delta K$) are not very large. In the opposite case, a stable soliton of one lattice emits a significant amount of radiation as approaches the interface in order to transform to a stable soliton of the other lattice (or a surface mode) resulting in dissipative dynamical behavior.\

Following the standard effective particle approach \cite{AcMoNe_89} the motion of the center of mass $x_0$ is given by the simple Newtonian equation of motion
\begin{equation}
m\frac{d^2x_0}{dz^2}=-\frac{\partial V_{eff}(x_0)}{\partial x_0} \label{Newton}
\end{equation}
with $m=\int|u|^2dx$ is the effective soliton mass and the effective potential $V_{eff}$ is given by
\begin{equation}
V_{eff}(x_0)=2\int_{-\infty}^{+\infty}n_0(x)|u(x-x_0)|^2dx \label{Veff}
\end{equation}
where for the evaluation of the effective potential one can use the solution of the unperturbed ($\epsilon=0$) NLSE: $u(z,x)=\eta \mbox{sech}\left[\eta(x-x_0)\right]e^{ivx/2+2\sigma}$, with $dx_0/dz=v$ and $d\sigma/dz=-v^2/8+\eta^2/2$. The effective particle approach is valid when the initial wave profile (in the form of a soliton) does not break up into multiple solitons and/or radiation, and this is the case we are interested in this work. The resulting effective potential depends strongly on the relation between three characteristic spatial scales, namely, the soliton width ($\sim \eta^{-1}$), the lattice period ($\sim (K^{\pm})^{-1}$) and the smoothness of the properties variation at the interface ($\sim a^{-1}$). The general form of the linear refractive index profile $n_0(x)$ (\ref{n0}) does not lead to an analytical derivation of the respective effective potential from Eq. (\ref{Veff}). However, the limiting form of $V_{eff}$ far from the interface (at $x_0 \rightarrow \pm\infty$) can be obtained analytically and it is identical to the effective potential of a single lattice
\begin{equation}
\lim_{x_0 \rightarrow \pm\infty}V_{eff}(x_0)=V^{(\pm)}\sin(K^{(\pm)}x_0+\phi) \label{Vasymptotic}
\end{equation}
with $V^{(\pm)}=\epsilon A^{(\pm)}(2\pi K^{(\pm)})/\sinh\left[\pi K^{(\pm)}/(2\eta)\right]$.
The properties of the formation and the dynamics of solitons can be directly obtained from the constant Hamiltonian (energy) of the equation of motion (\ref{Newton}) $H=mv^2/2+V_{eff}(x_0)$ from which the phase space of the respective system can be readily obtained.\

Far from the interface, similarly to the case of a single lattice, stationary LS are formed at the extrema of the potential  with LS with $x_0$ at the minima (maxima) of the potential corresponding to stable (unstable) modes, for solitons with any $\eta$ \cite{PeSuKi_04,KoHi_08}. While the asymptotic form of the potential is identical for all $\eta$, its depth $V^{(\pm)}$ depends strongly on the ratio $K^{(\pm)}/\eta$. This depth determines the critical value of the soliton velocity $v_c=(2V^{(\pm)}/m)^{1/2}$ above which a stable LS can be de-trapped from the potential minimum and move across the lattice. \

On the other hand, close to the interface not only the strength but also the form of the potential depends strongly on the characteristics of the soliton, namely on $\eta$. A variety of qualitatively different forms of the effective potentials can occur, in the same structure for different solitons, as we show in the following. The number and the position of local minima and maxima of the effective potential provide important information for both the statics and the dynamics of solitons in such structures, that is the location and stability of stationary SS as well as the transmission and reflection of mobile LS. \  
\begin{figure}[h]
    \begin{center}
        \subfigure{\scalebox{0.12}{\includegraphics{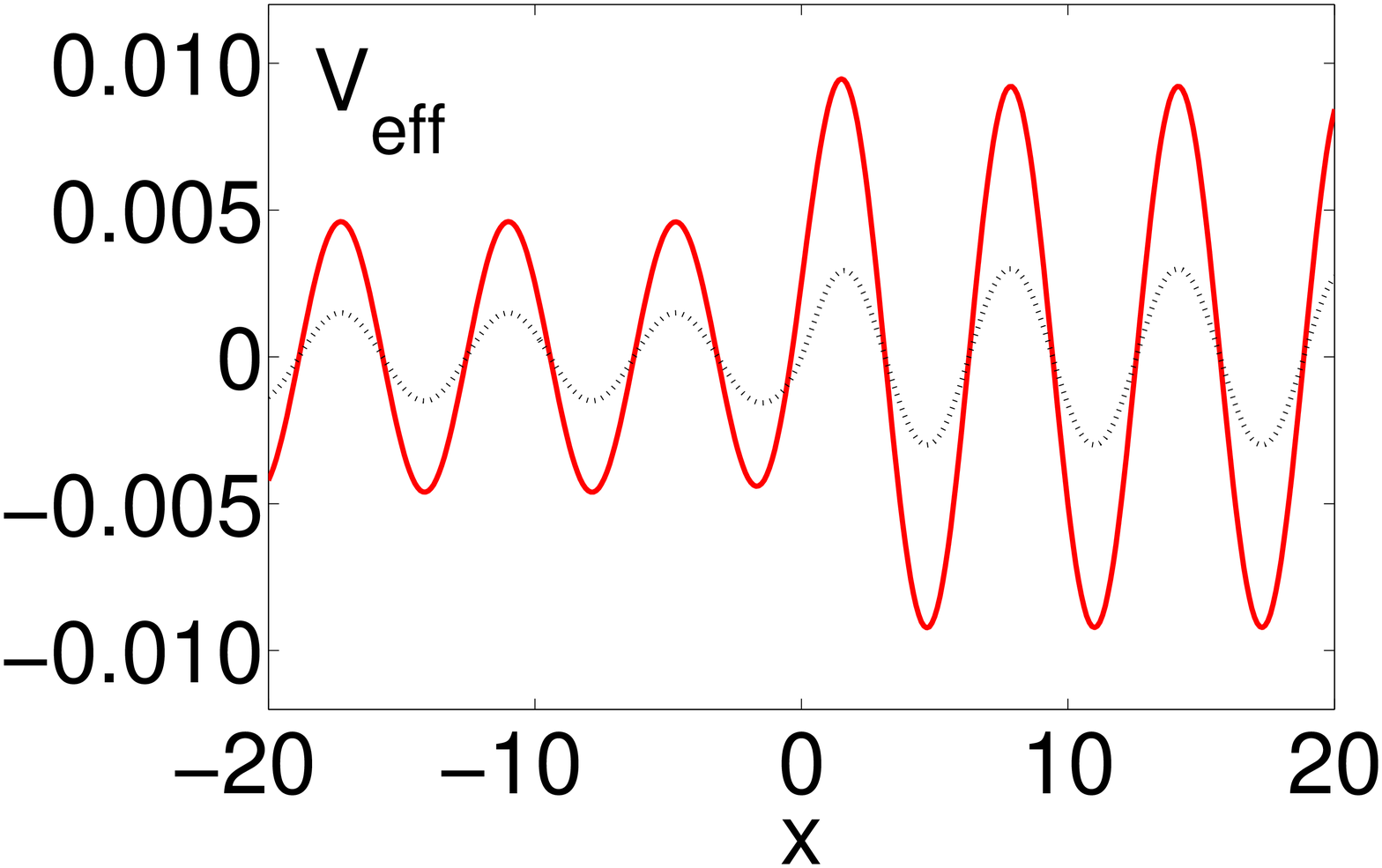}}}
        \subfigure{\scalebox{0.12}{\includegraphics{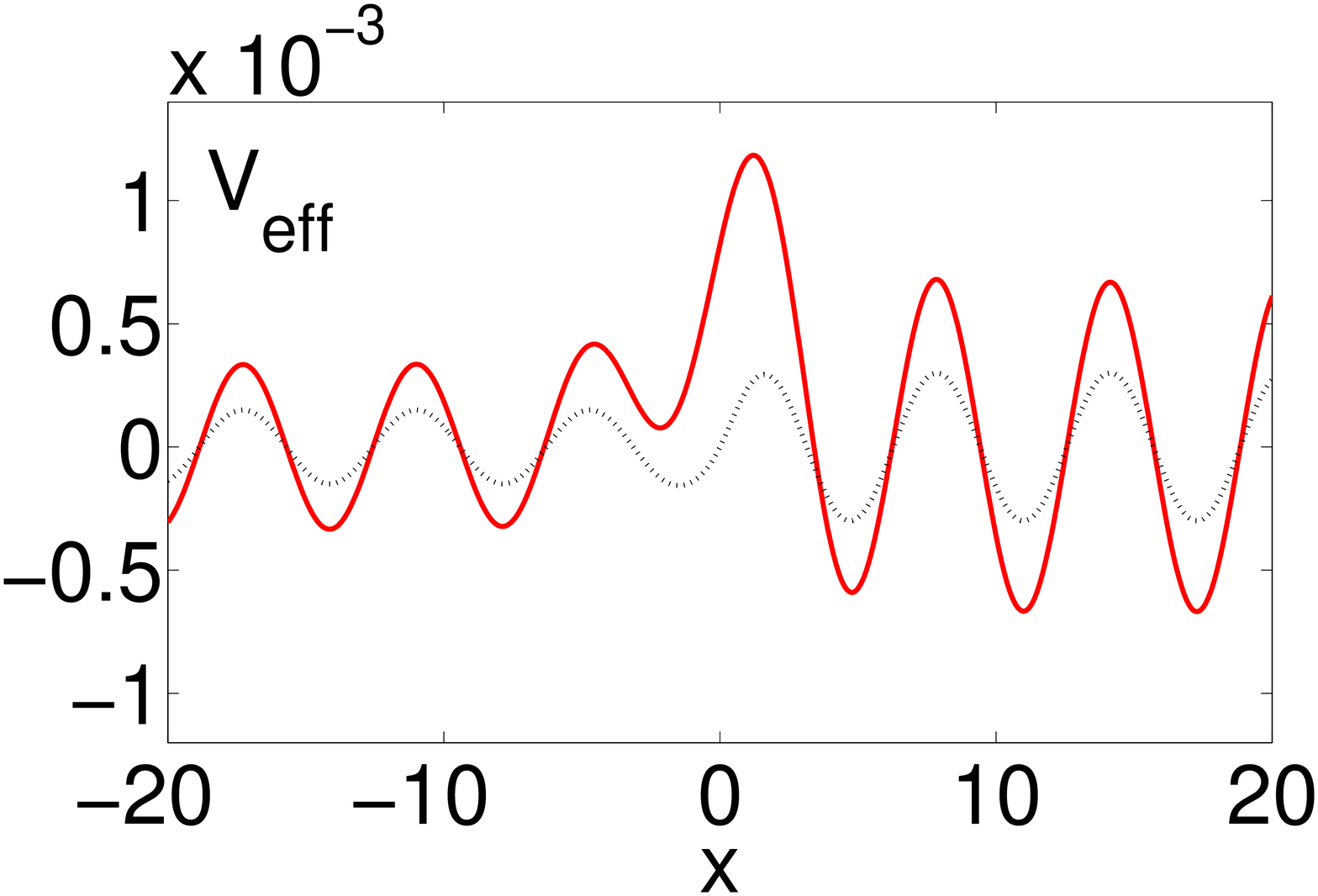}}}\\
        \subfigure{\scalebox{0.12}{\includegraphics{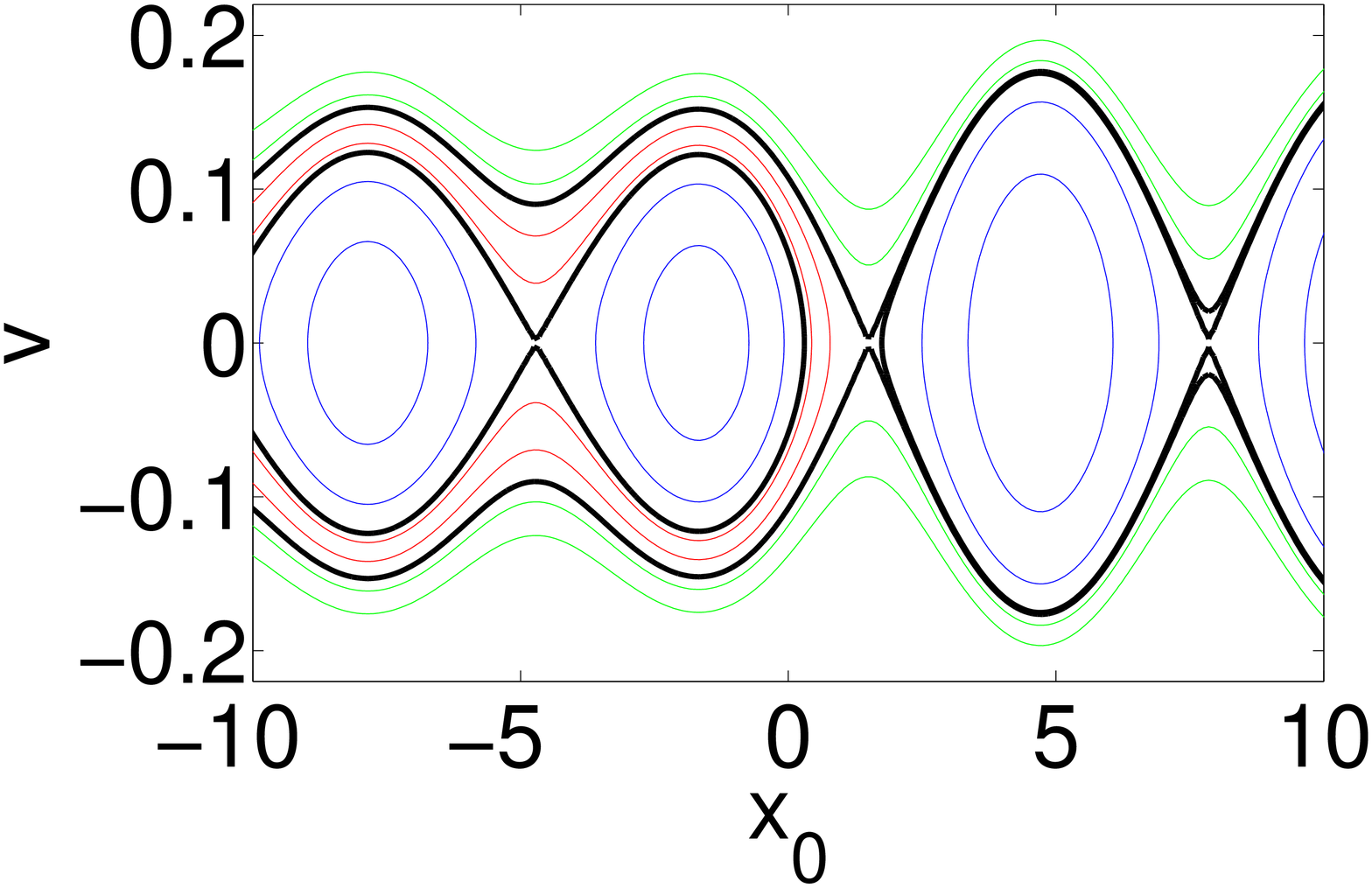}}}
        \subfigure{\scalebox{0.12}{\includegraphics{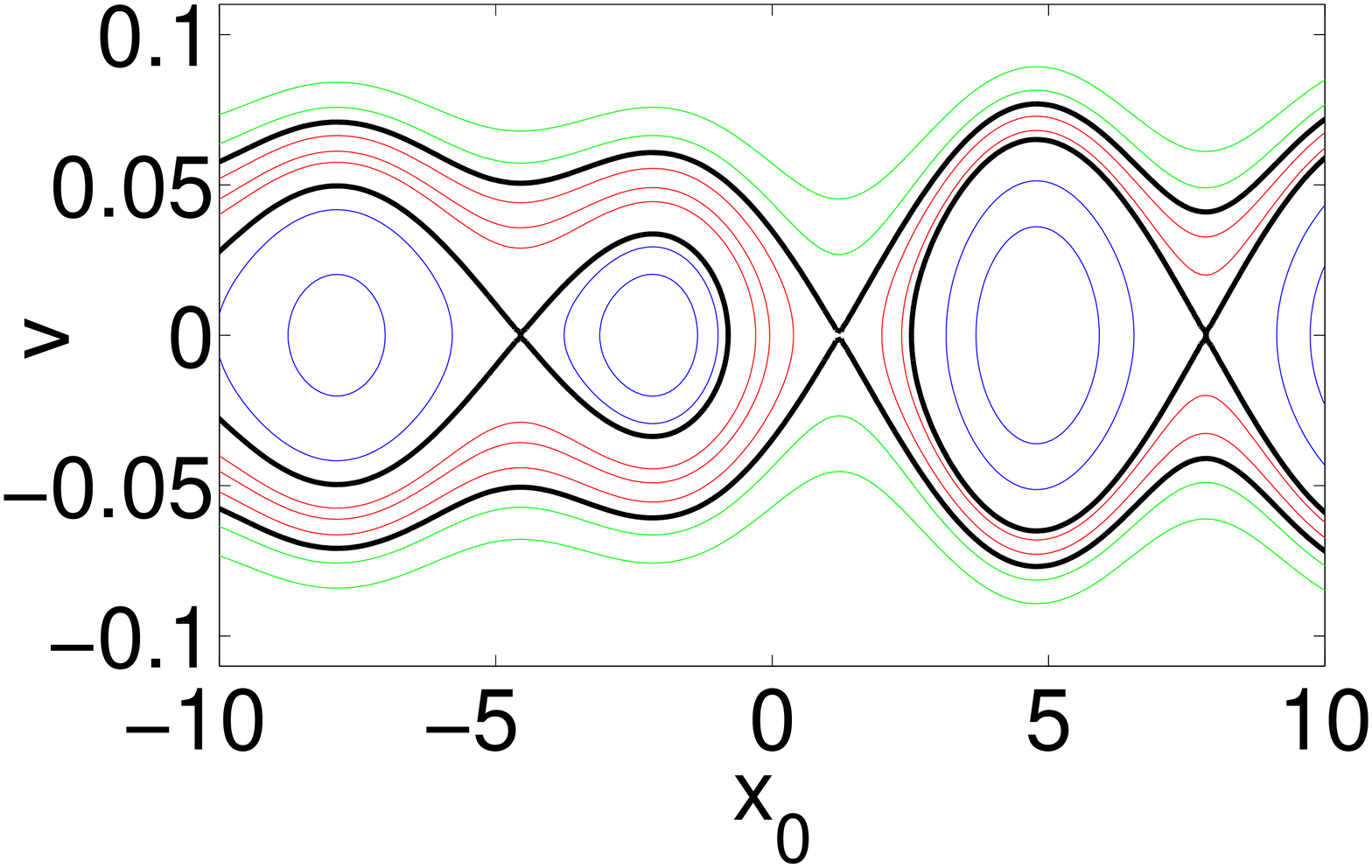}}}
        \caption{(Top) Effective potential (red online) and $n_0(x)$ (dotted line, out of scale), and (Bottom) phase plane  for an effective particle corresponding to a soliton with $\eta=0.6$ (Left), $\eta=0.3$ (Right). Trapped (blue online), reflected (red online) and transmitted (green online) orbits are shown. The underlying structure parameters are $\epsilon=0.01$, $A^{(-)}=0.5$, $A^{(+)}=0.5$, $K^{(-)}=K^{(+)}=1$, $a=1$, $\phi=0$. }
    \end{center}
\end{figure}

\begin{figure}[h]
    \begin{center}
        \subfigure{\scalebox{0.10}{\includegraphics{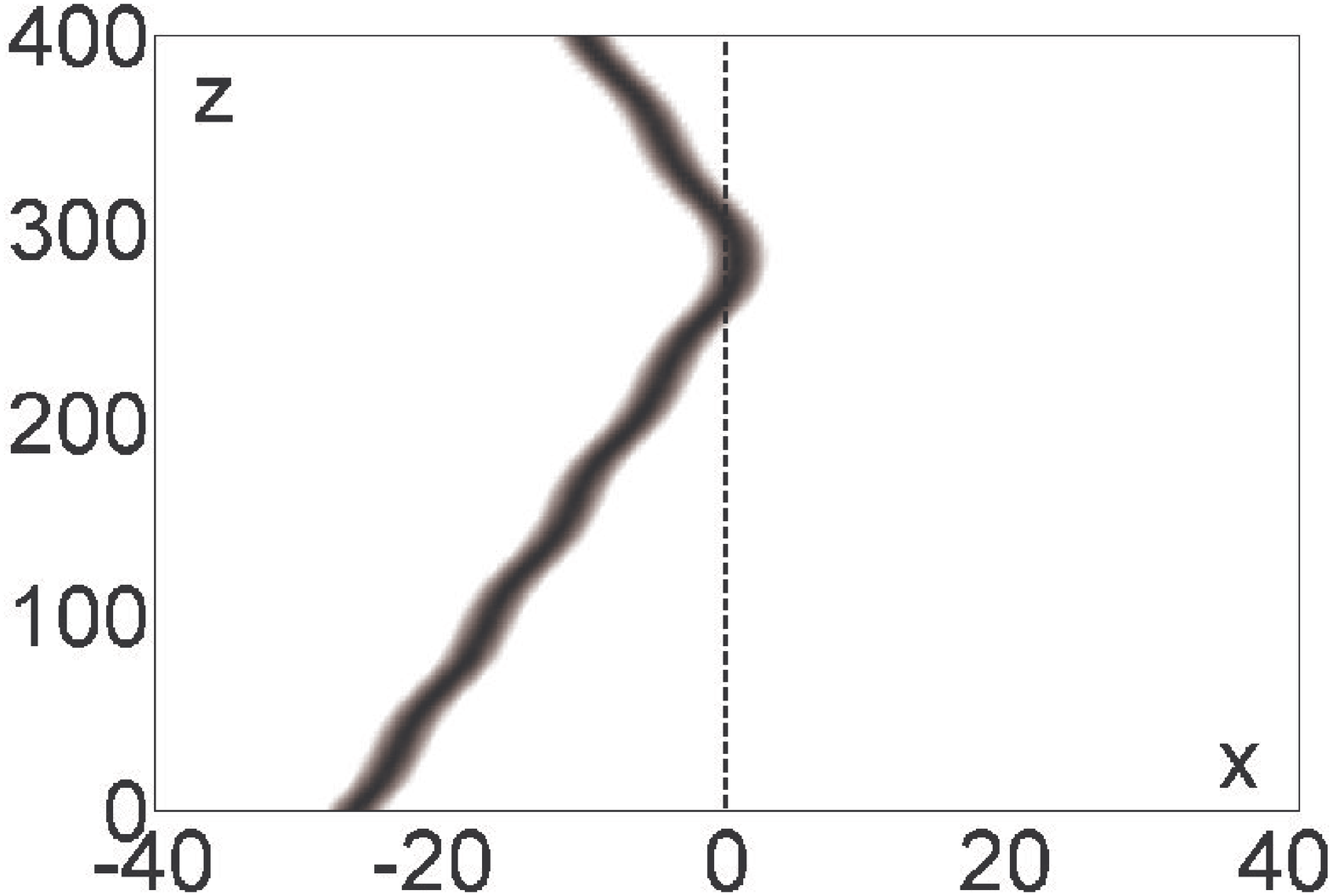}}}
        \subfigure{\scalebox{0.10}{\includegraphics{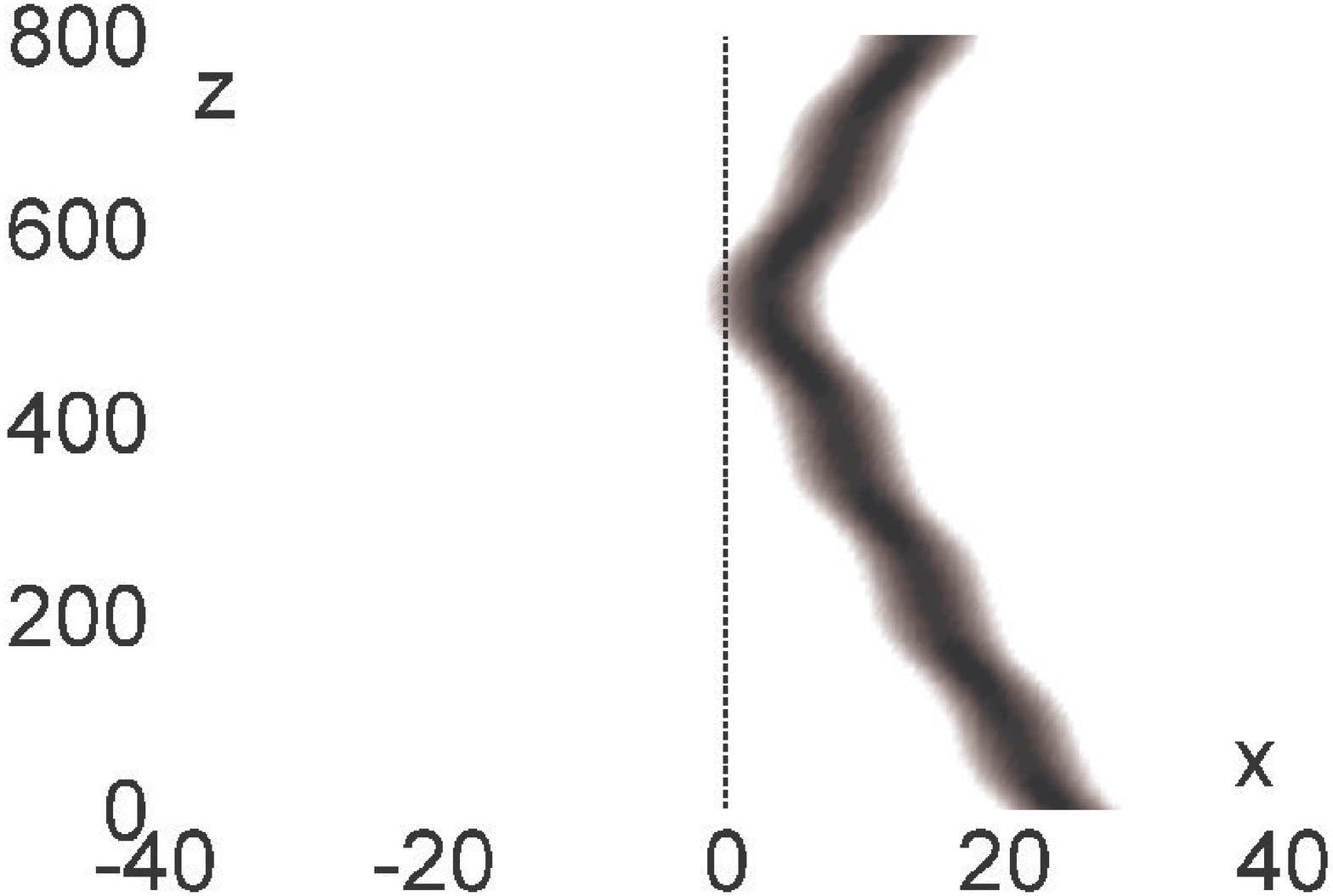}}}\\
        \subfigure{\scalebox{0.10}{\includegraphics{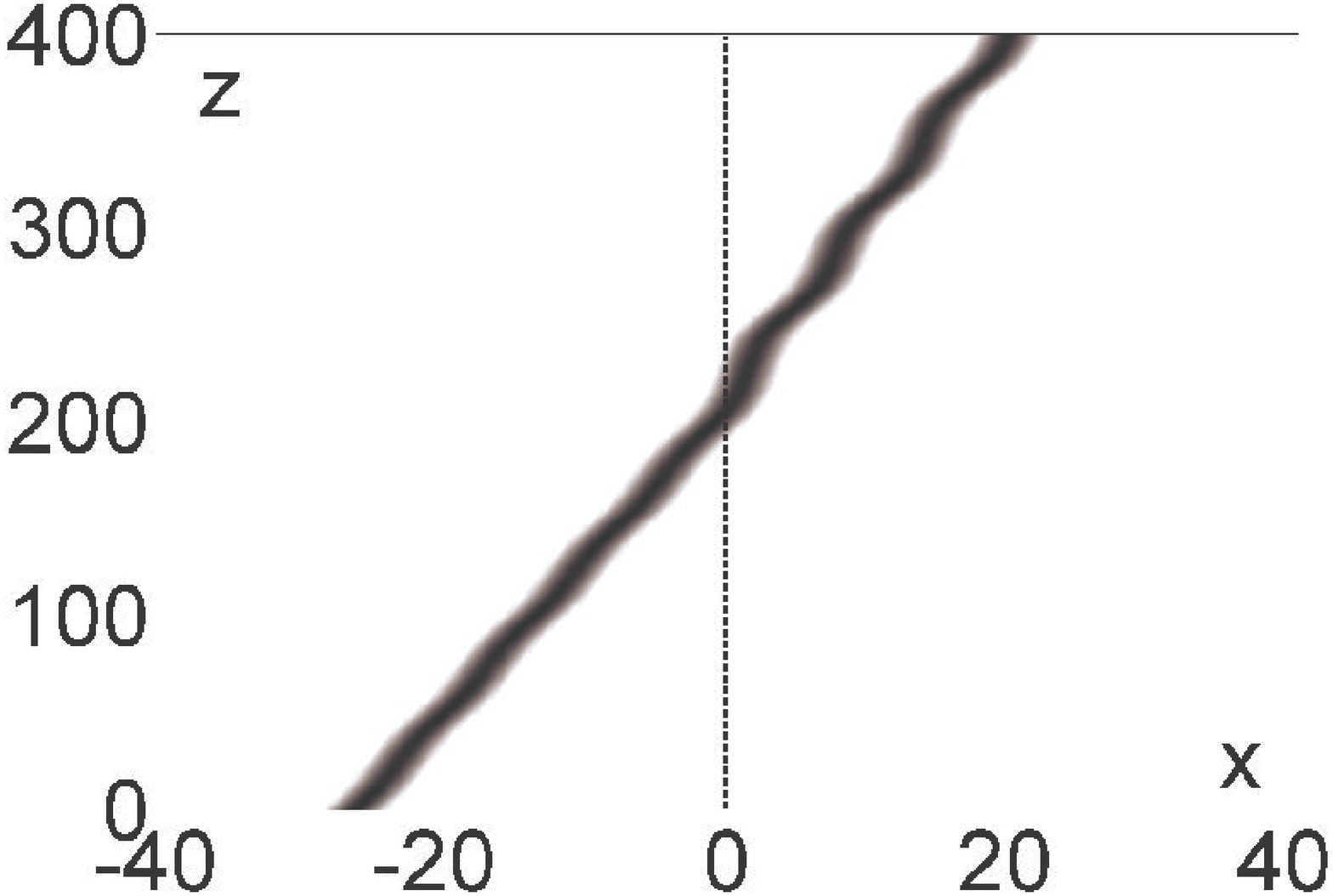}}}
        \subfigure{\scalebox{0.10}{\includegraphics{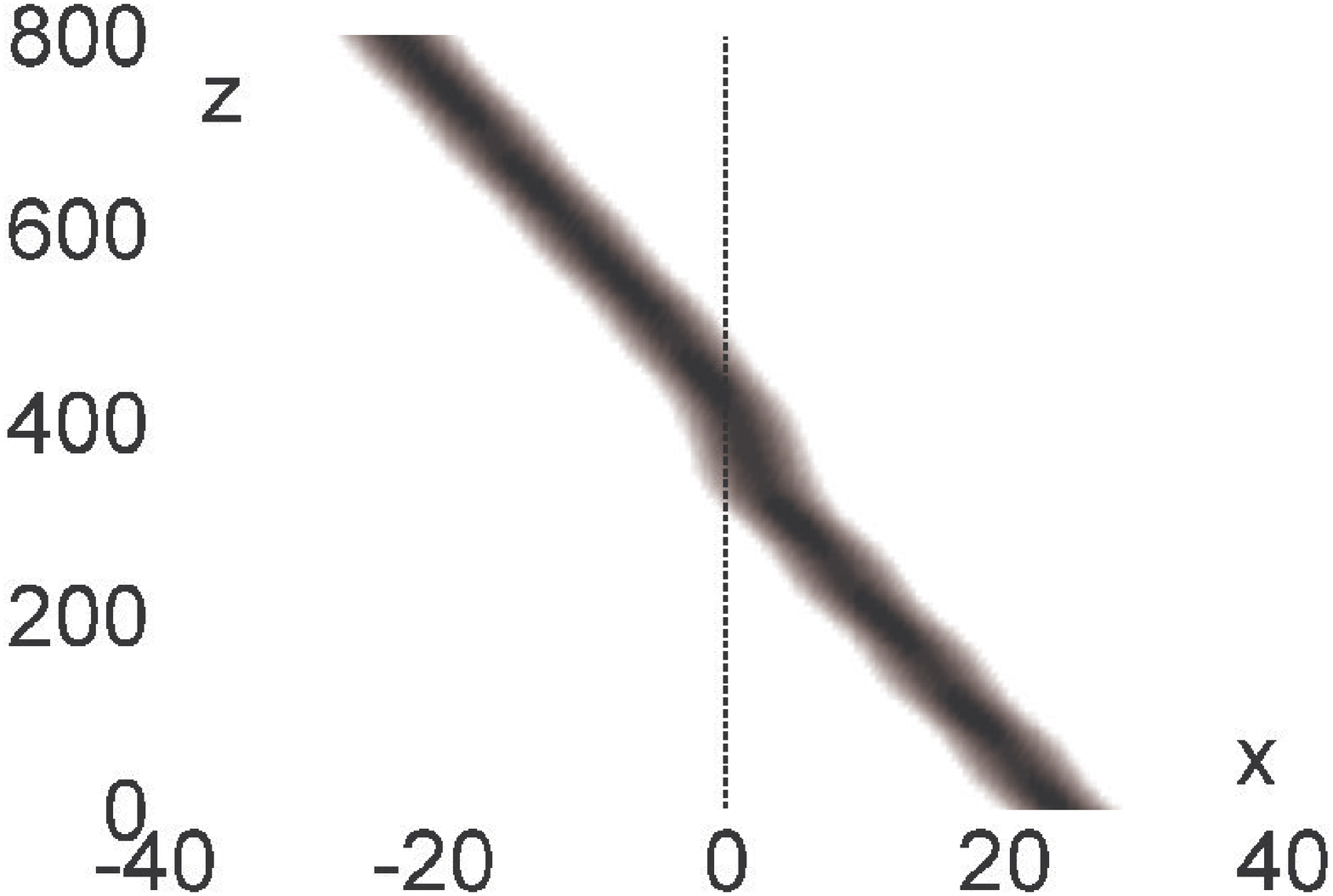}}}
        \caption{Propagation of a LS with $\eta=0.6$ (Left), $\eta=0.3$ (Right) with initial position $x_0=-17\pi/2$ (Left) , $x_0=15\pi/2$ (Right), corresponding to minimum of $V_{eff}$, and initial velocity $v=-14 \times 10^{-2}$ (Left, Top), $v=-16 \times 10^{-2}$ (Left, Bottom), $v=7.2 \times 10^{-2}$ (Right, Top), $v=8.4 \times 10^{-2}$ (Right, Bottom). The parameters of the structure are as in Fig. 1. }
    \end{center}
\end{figure}
Firstly, we consider a structure where the interfaced lattices have the same wavenumber $K^{(-)}=K^{(+)}=1$ and different amplitudes $A^{(-)}=0.5, A^{(+)}=1$. In all the following examples, we take $\epsilon=0.01$ in order to ensure validity of the effective particle approach, and we consider a smooth property variation having $a=1$. For the case where $\phi=0$, the effective potential is shown in Figs. 1(top) for two solitons of different power (and width) having $\eta=0.6$ and $\eta=0.3$. The respective phase plane diagrams $(x_0,v)$ can be directly obtained as in Figs. 1(bottom), so that both soliton statics, stability and dynamics can be determined. This diagram plays the role of a nonlinear analogue of Snell's law, describing LS transmission and reflection. Stable (unstable) LS or SS correspond to minima (maxima) of the effective potential. Close to the stable stationary LS or SS, trapped states with $v\neq 0$ can oscillate in the respective potential well. While stationary LS formed quite far from the interface are trivially located in the refractive index minima (stable) or maxima (unstable) for every soliton width $\eta$, the situation is drastically different in the area close to the interface. For $\eta=0.6$ [Fig. 1(left)], the pattern of the effective potential is similar to the underlying linear refractive index profile. LS reflection can occur for appropriate $v$ only for waves incident from left to right with initial $v$ larger $v_c^{(-)}$ but lower than $v_c^{(+)}$, while LS with initial velocities larger than $v_c^{(+)}$ are transmitted through the interface [Figs. 2(left)]. The situation is qualitatively different for solitons having $\eta=0.3$ [Fig. 1(right)], where an effective potential barrier, associated to an unstable SS, can additionally lead to reflection of waves incident from the right hand side [Figs. 2(right)]. LS transmission through the interface requires higher $v$ due to the interface induced barrier. The formation of this potential barrier cannot be attributed to either of the two lattices but to their interfacing, and it does not have a uniform effect on all solitons.\
\begin{figure}[h]
    \begin{center}
        \subfigure{\scalebox{0.12}{\includegraphics{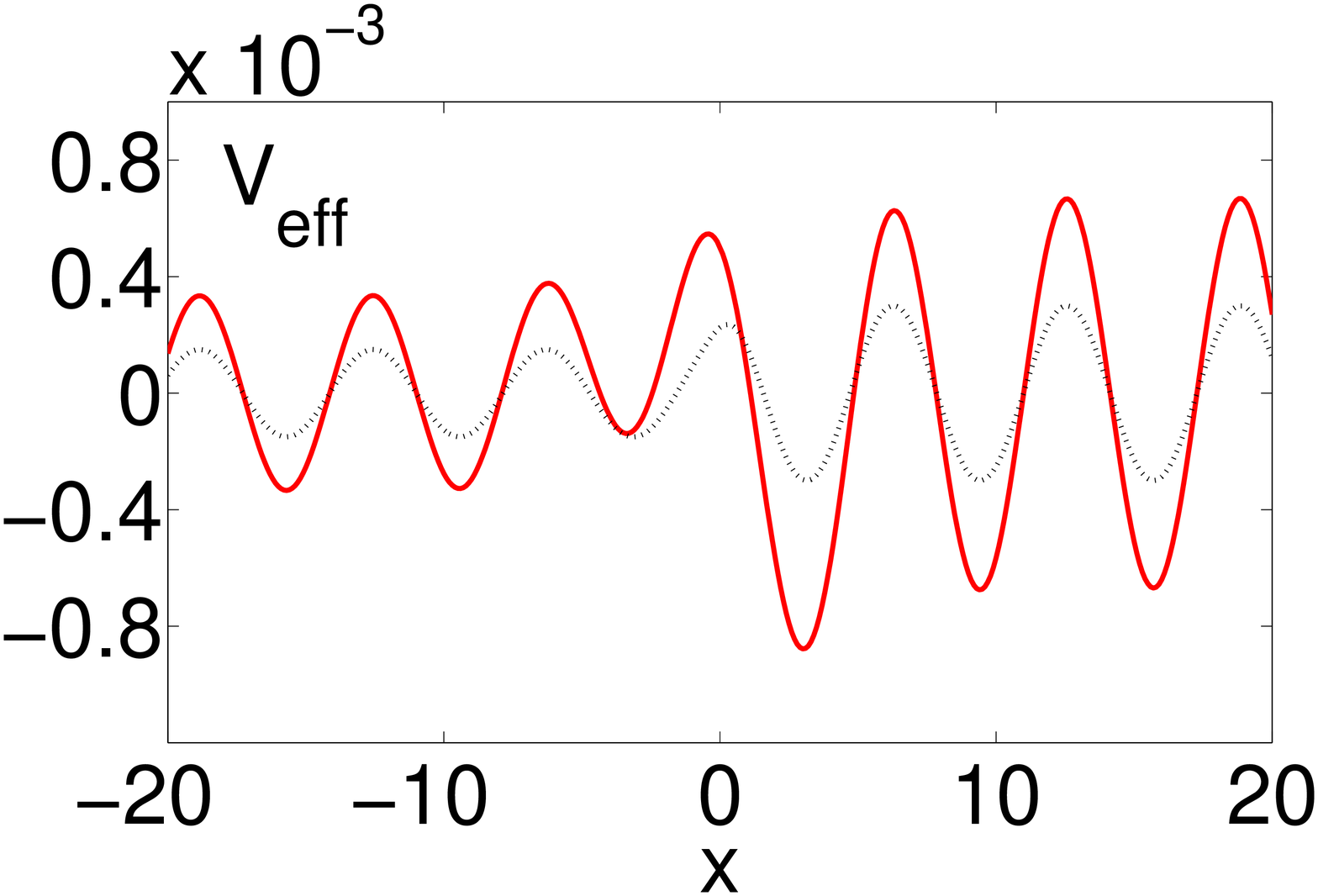}}}
        \subfigure{\scalebox{0.12}{\includegraphics{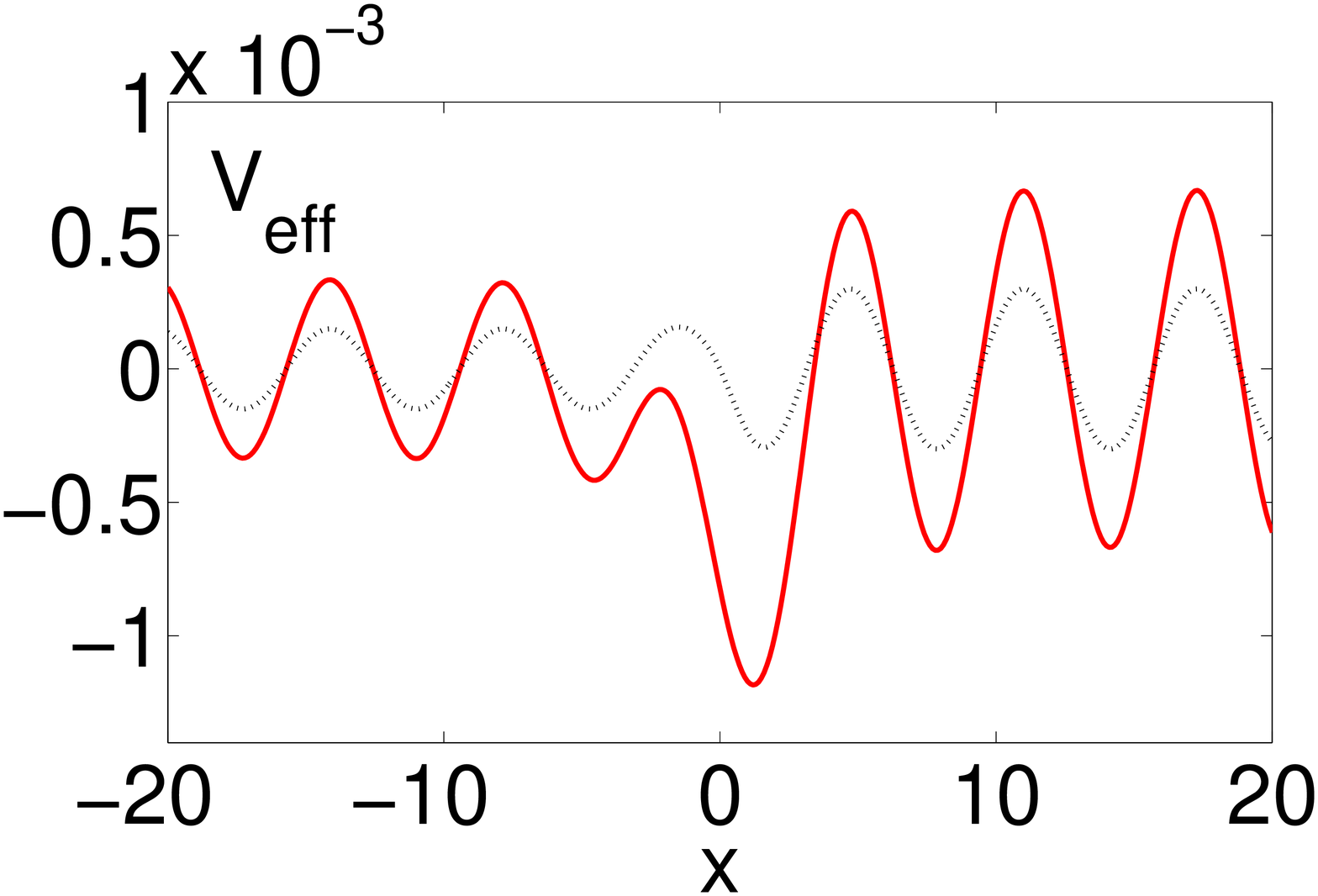}}}
        \caption{Effective potential (red online) and $n_0(x)$ (dotted line, out of scale) for a LS having $\eta=0.3$ in a structure with parameters as in Fig. 1, except that $\phi=\pi/2$ (Left), $\phi=\pi$ (Right).}
    \end{center}
\end{figure}

\begin{figure}[h]
    \begin{center}
        \subfigure{\scalebox{0.10}{\includegraphics{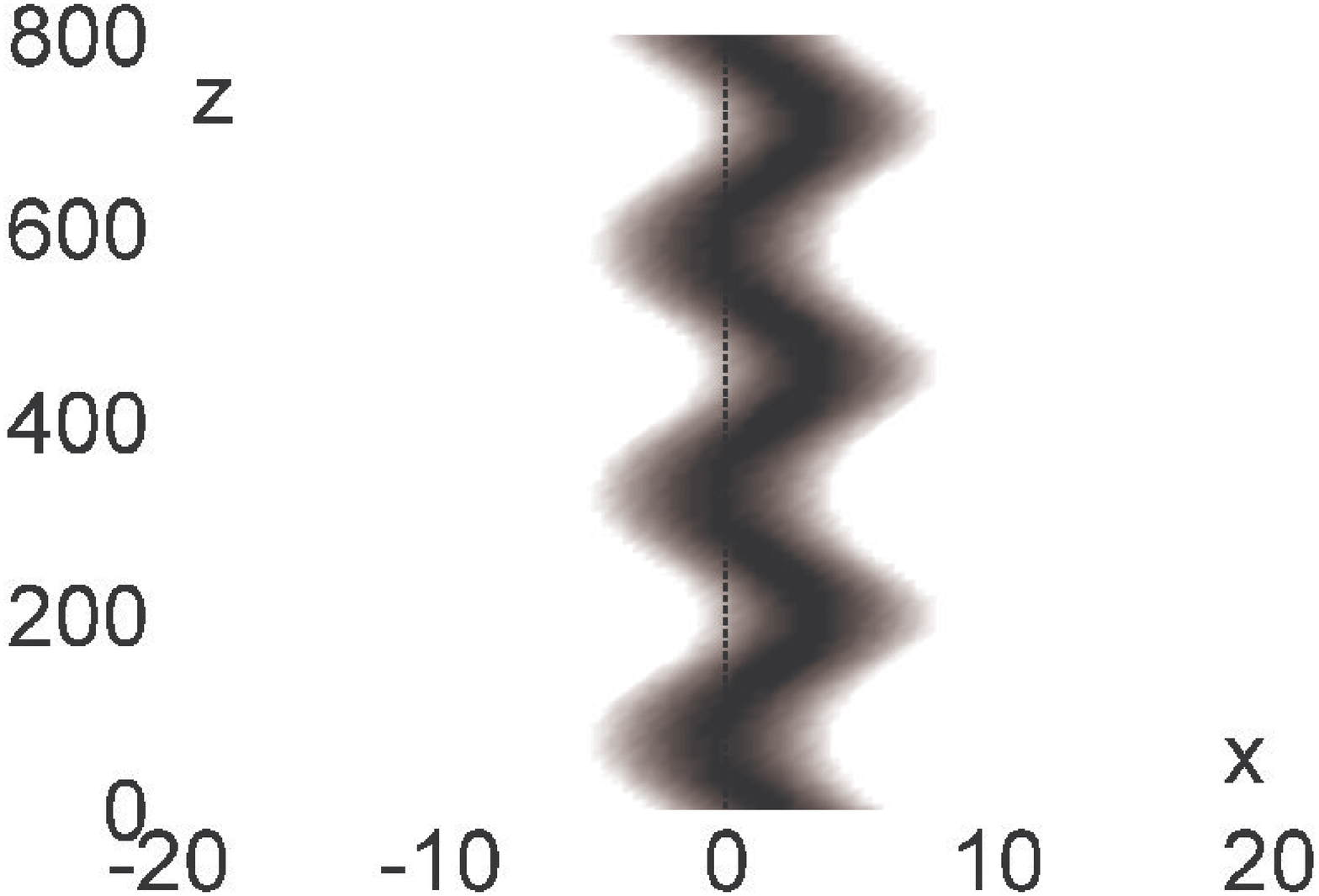}}}
        \subfigure{\scalebox{0.10}{\includegraphics{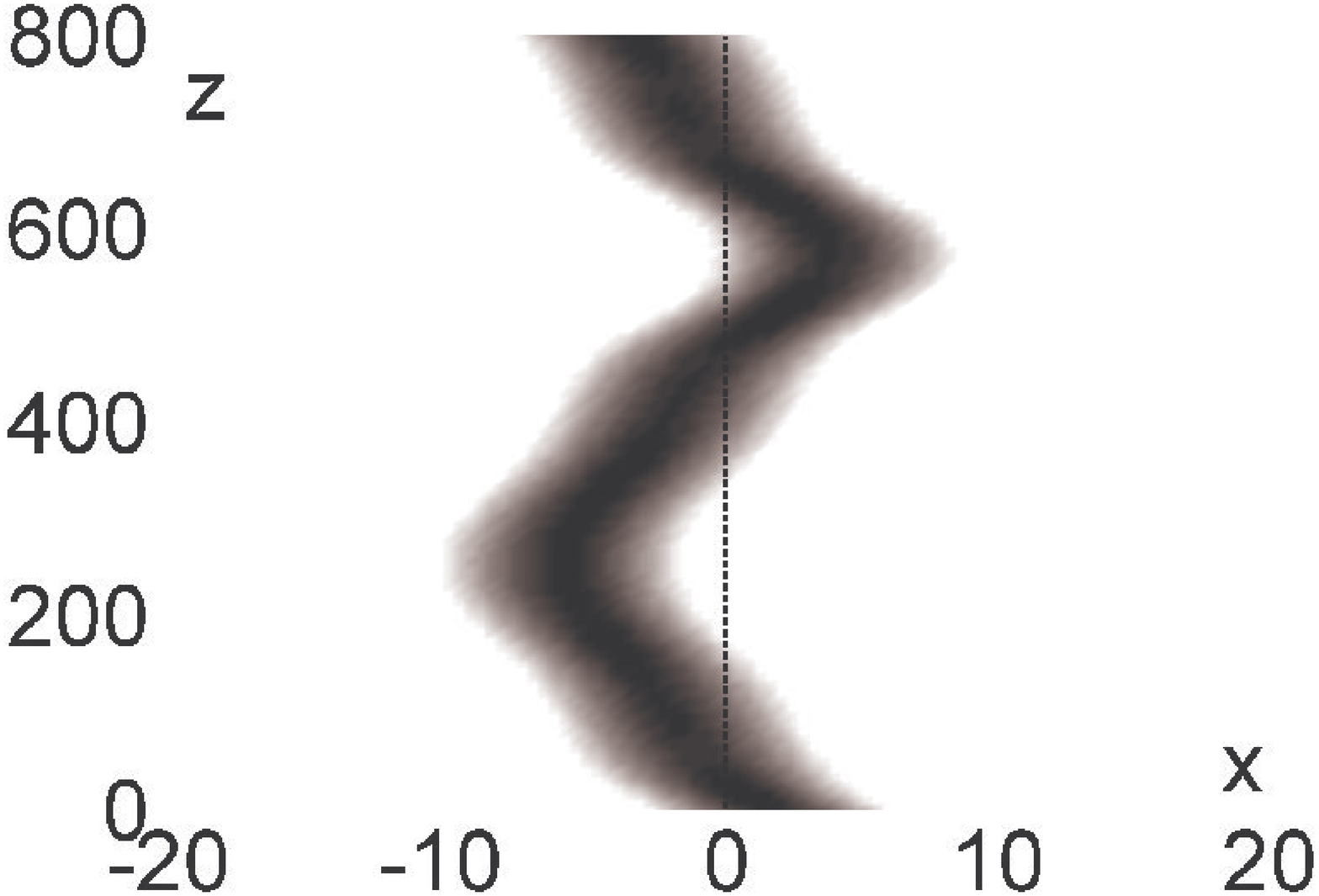}}}
        \caption{Propagation of SS with $\eta=0.3$, initial position $x_0=1.2$ [global minimum of $V_{eff}$ in Fig. 3(left)] and initial velocity $v=5.4 \times 10^{(-2)}$ (Left), $v=6.6 \times 10^{(-2)}$ (Right).}
    \end{center}
\end{figure}
The constant phase $\phi$ [related to the amplitude of the linear refractive index at the interface (at $x=0$)] plays a nontrivial role in the form of the effective potential and the corresponding soliton statics and dynamics in the structure. For $\phi=\pi/2$ [Fig. 3(left)] and for solitons with $\eta=0.3$, we have two local minima of the effective potential in both sides of the origin corresponding to stable SS. The case of $\phi=\pi$ and a soliton with $\eta=0.3$ [Fig. 3(left)], results to an effective potential of opposite sign with the one shown in Fig. 1(top, left). The global minimum of the potential close to the interface corresponds to a stable SS. Solitons with initial velocities, corresponding to a total energy level inside the potential well are trapped and oscillate around the position of the potential minimum \cite{StSmRu_07}. The corresponding de-trapping velocities can be directly obtained from the effective potential. This is an important advantage of the effective particle approach because, in  addition to determining the stability of a stationary SS (as linear stability analysis could also do), it provides a measure of the degree of stability of the SS in terms of the depth of the effective potential and the de-trapping velocity. The latter is directly related to the launching angle of the soliton at $z=0$, and gives the tolerance on the alignment precision for soliton trapping. In Fig. 4(left), we show a trapped SS with initial energy less than the local maximum located on the left of origin, oscillating around the position of the potential minimum. For velocities corresponding to initial energies larger than the local potential maximum on the left of the origin [Fig. 3(left)], the amplitude of the soliton position oscillation is larger [Fig. 4(left)], as predicted from the effective potential. \

\begin{figure}[h]
    \begin{center}
        \subfigure{\scalebox{0.12}{\includegraphics{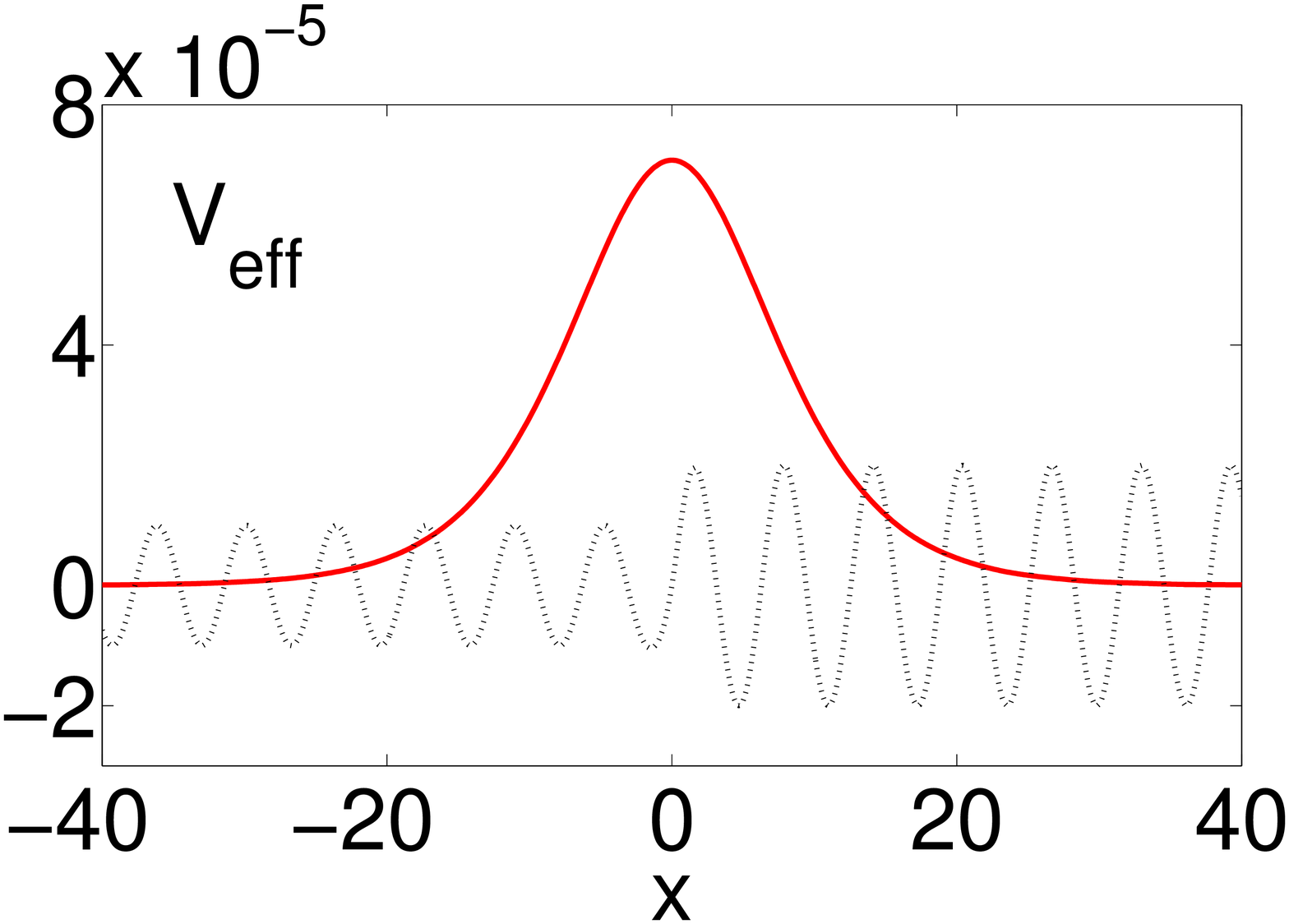}}}
        \subfigure{\scalebox{0.12}{\includegraphics{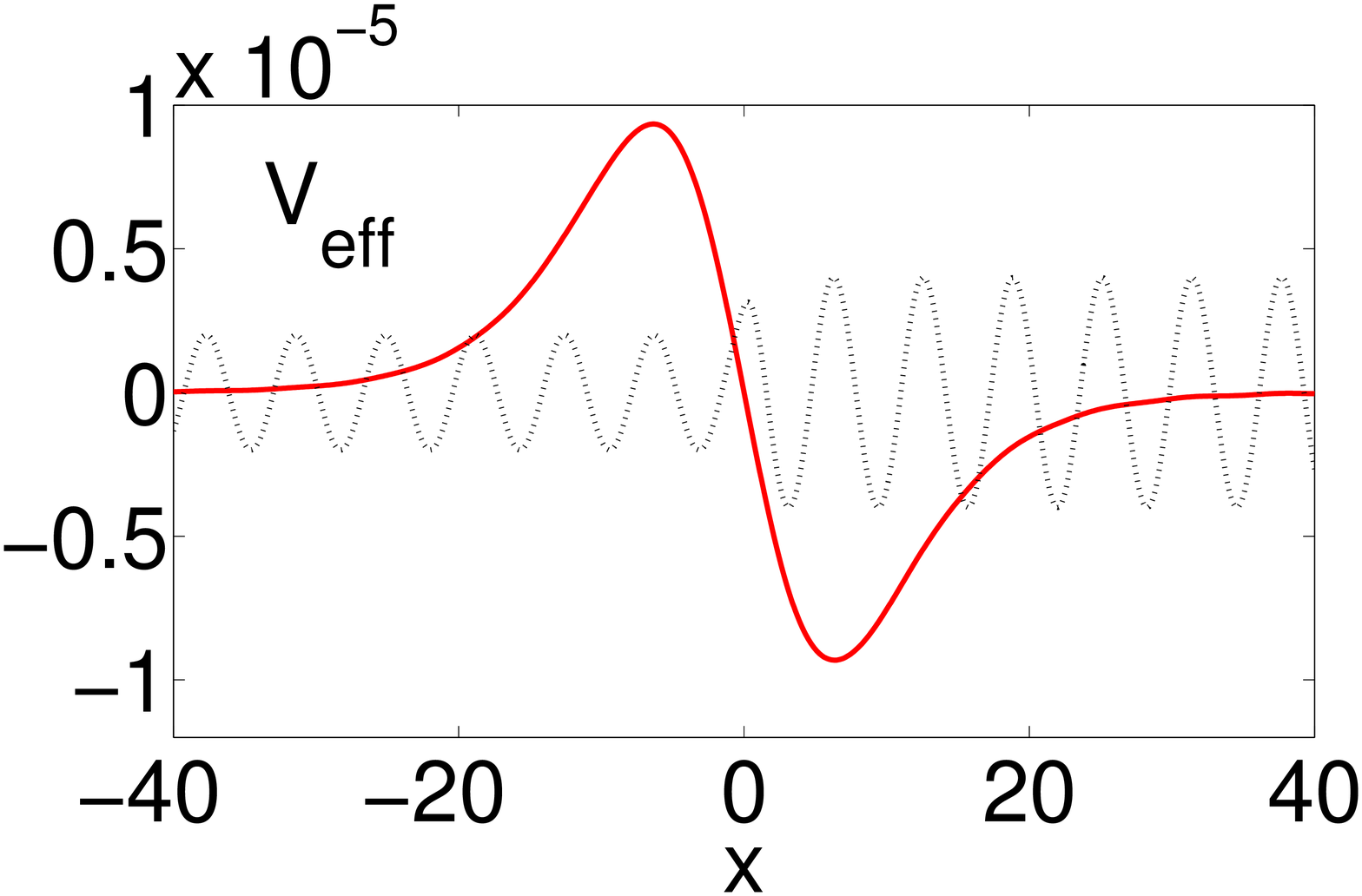}}}
        \caption{Effective potential (red online) and $n_0(x)$ (dotted line, out of scale) for a wide LS with $\eta=0.1$. The underlying structure parameters are as in previous Figs. except that $\phi=0$ (Left), $\phi=\pi/2$ (Right).}
    \end{center}
\end{figure}

From the previous cases it is shown that the formation of nontrivial effective potential patterns including barriers and wells, in the vicinity of the interface, is mostly related to solitons with large (in comparison to the lattice period) widths ($\eta$). It is remarkable that for wide LS, although the amplitude of the effective potential is very small sufficiently far from the interface [as obtained by Eq. (\ref{Vasymptotic})], it has large values close to the interface. This means that although a wide LS does not "feel" the lattice, it does "feel" the interface. Such cases, for a LS with $\eta=0.1$, are shown in Figs. 5 for $\phi=0$(left) and $\phi=\pi/2$. According to the respective effective potentials, a LS which is untrapped from the lattice far from the interface and can move, it can be reflected from the interface even if it has a large $v$  [Fig. 5(left)]. Also, stable and trapped SS can be formed in the vicinity of the interface in a position depending on the phase constant $\phi$ [Fig. 5(right)].\
\begin{figure}[h]
    \begin{center}
        \subfigure{\scalebox{0.12}{\includegraphics{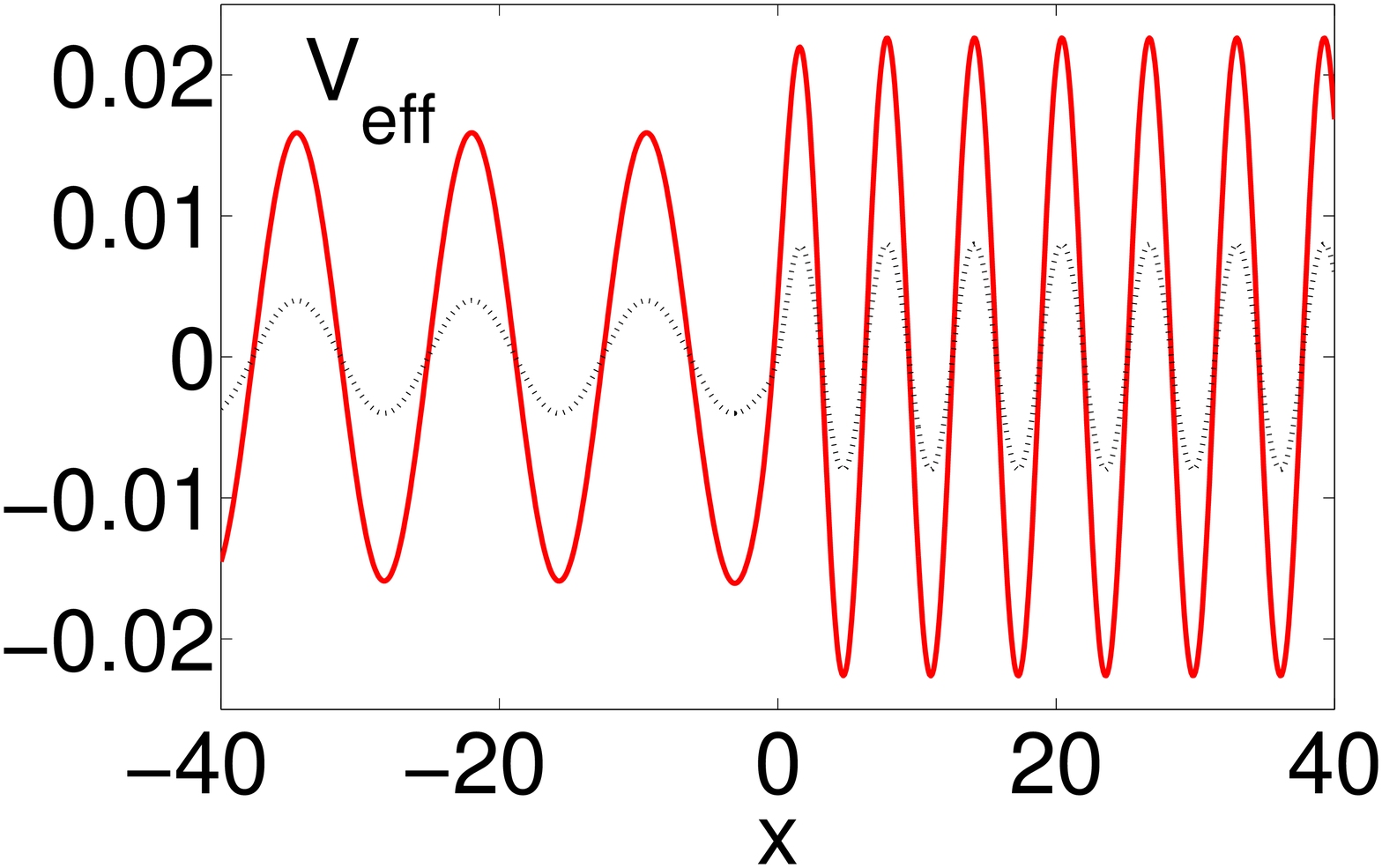}}}
        \subfigure{\scalebox{0.12}{\includegraphics{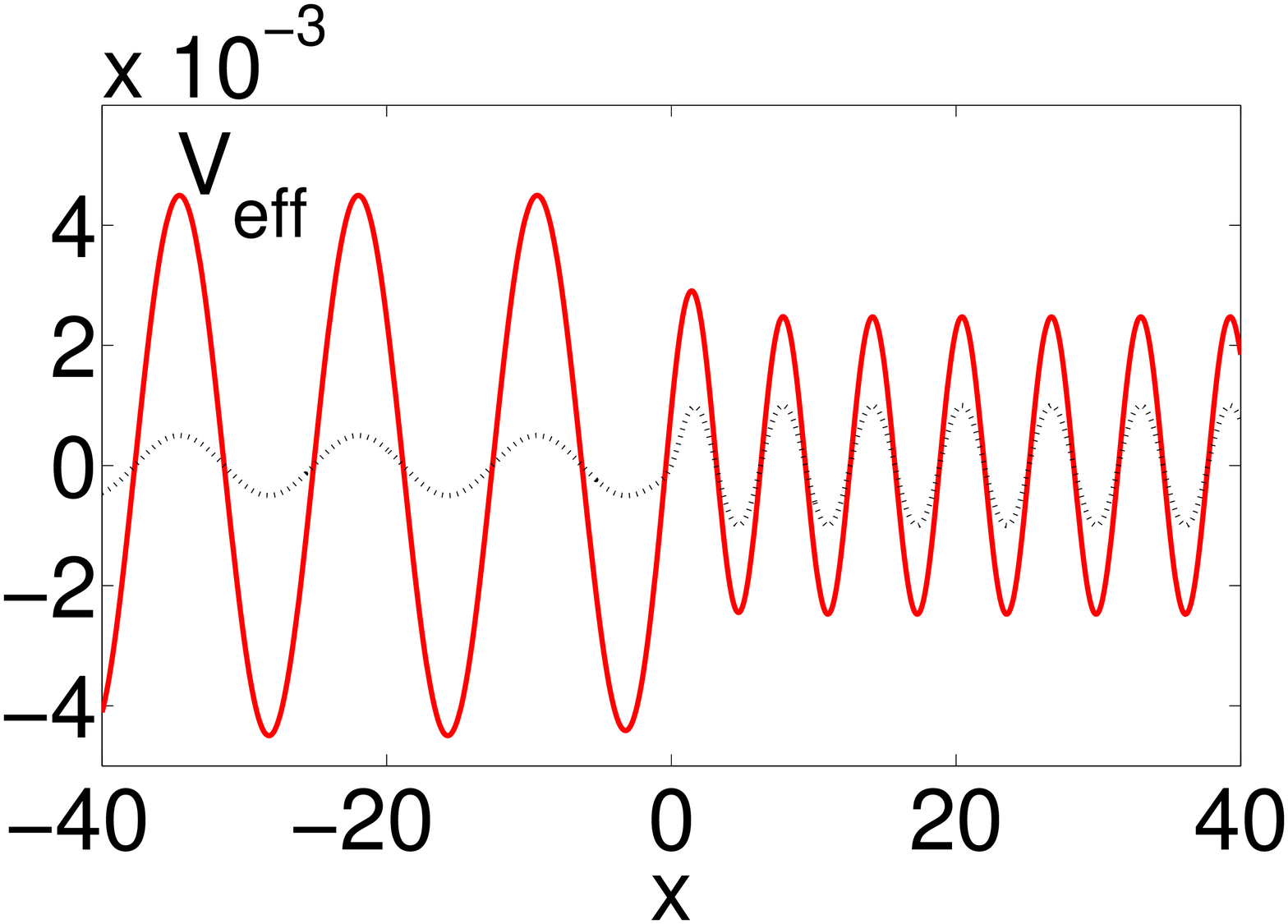}}}
        \caption{Effective potential (red online) and $n_0(x)$ (dotted line, out of scale) for LS having $\eta=0.9$ (Left), $\eta=0.4$ (Right) in a structure consisting of two interfaced lattices with different periods. Parameters: $\epsilon=0.01$, $A^{(-)}=0.5$, $A^{(+)}=0.5$, $K^{(-)}=0.5$, $K^{(+)}=1$, $a=1$, $\phi=0$.}
    \end{center}
\end{figure}

\begin{figure}[h]
    \begin{center}
        \subfigure{\scalebox{0.10}{\includegraphics{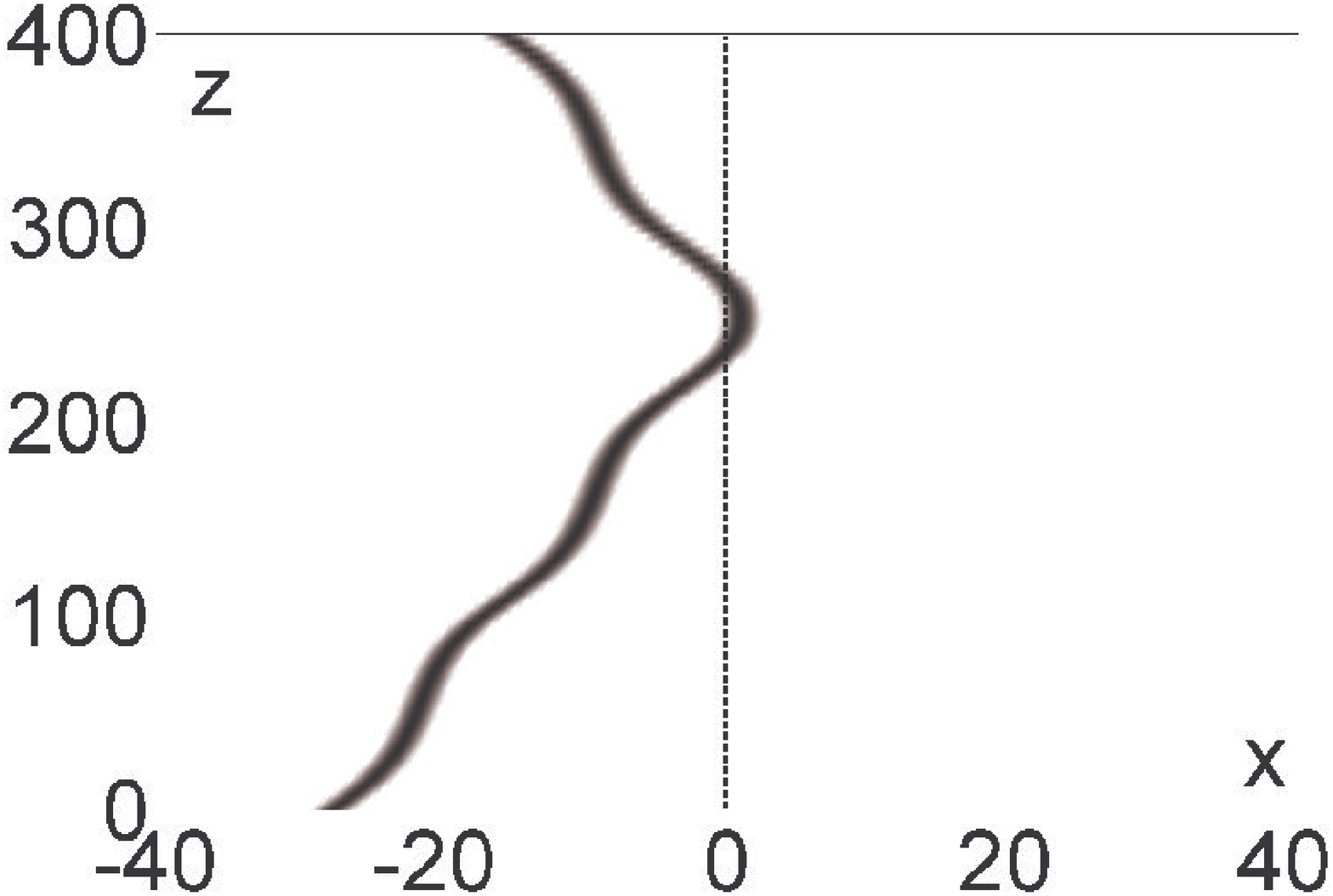}}}
        \subfigure{\scalebox{0.10}{\includegraphics{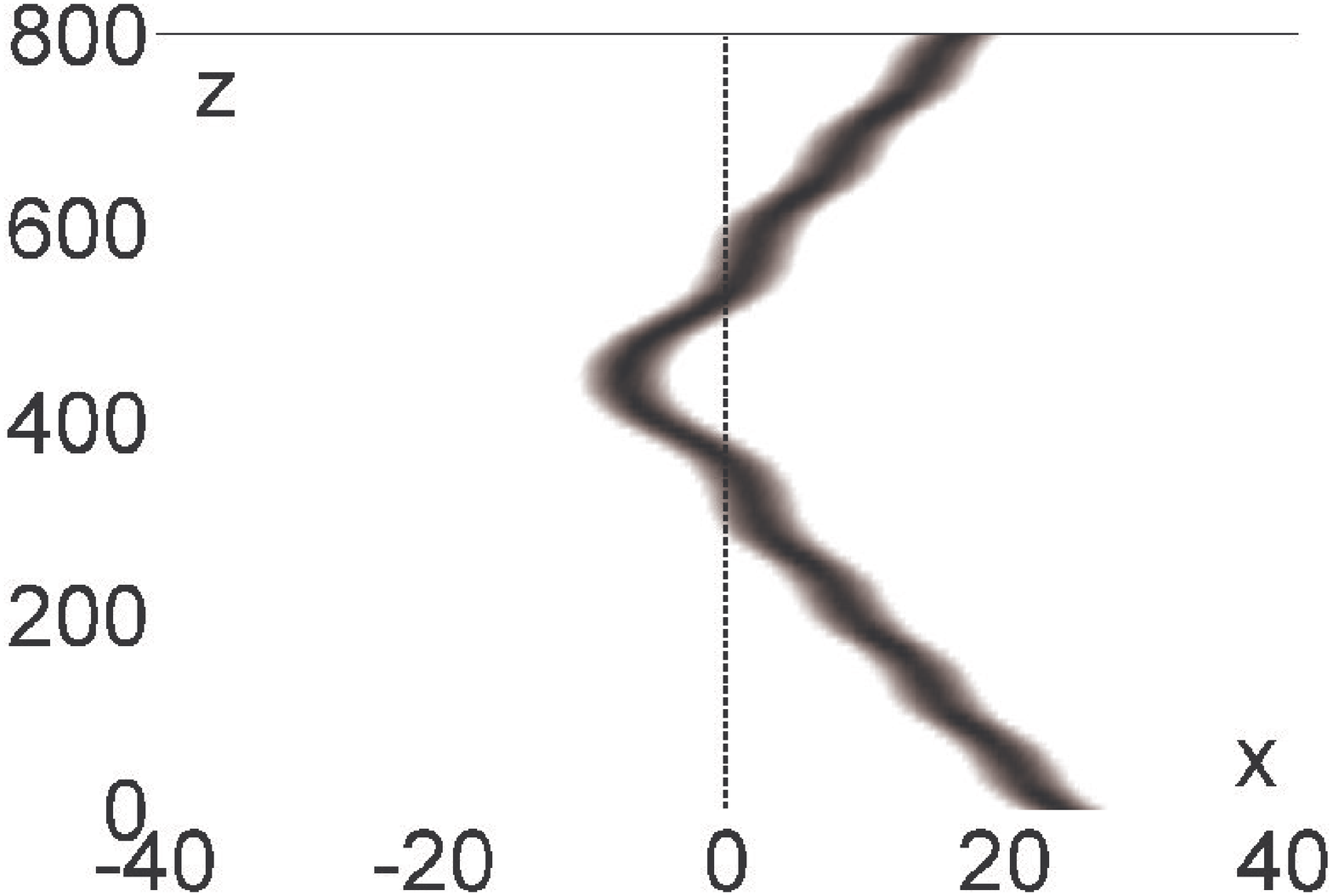}}}
        \caption{Reflection of a LS with $\eta=0.9$ (Left), $\eta=0.4$ (Right) incident at the interface from an initial position $x_0=-18\pi/2$ (Left), $x_0=15\pi/2$ (Right) with $v=-19.6 \times 10^{-2}$ (Left), $v=12 \times 10^{-2}$ (Right).}
    \end{center}
\end{figure}
Finally, an interesting case occurs when not only the amplitudes of the two interfaced lattices are different, but also the wavenumbers $(K^{(\pm)})$. In Figs. 6, we show the corresponding effective potentials for solitons having $\eta=0.9$ (left) and $\eta=0.4$ (right) in a configuration where $A^{(-)}=0.5$, $A^{(+)}=1$, $K^{(-)}=0.5$ and $K^{(+)}=1$. It is interesting that the same structure results in a qualitatively different "landscape" for different solitons. Therefore, for solitons with $\eta=0.9$, $V^{(-)}<V^{(+)}$ and reflection can take place only for solitons incident from the left, while the opposite holds for solitons with $\eta=0.4$ (Figs. 7). For solitons with $\eta\simeq 0.6$ no reflection can take place. Therefore, a power-dependent directional effect is introduced.\

In the previous cases we have investigated, in terms of the effective potential (and the corresponding phase plane), whether a soliton with a given initial position and velocity can be trapped, reflected or transmitted in the interface between two dissimilar lattices. In the case of reflection, an additional information of interest is related to the position where a LS is actually reflected, which can be quite different under certain parameter selections as shown for example in Figs. 7. This difference is manifested as a displacement of the reflected wave from its geometric optics path. For the case of linear wavepackets, this effect is referred as the Goos-H\"{a}ncen shift. For the nonlinear case of LS soliton reflection at an interface and within the effective particle approximation the natural analogue of the Goos-H\"{a}ncen shift $\delta z$ is defined as the difference in the asymptotic value of the $z$ location of the effective particle as $x_0\rightarrow \pm \infty$ and the location of a free particle bouncing at the interface $x=0$ \cite{AcMoNe_89}. The respective analytical expression is $\delta z=\int_{\pm \infty}^{x_r}[v^{-1}(x)-v_0^{-1}]dx+x_r/v_0$, where $v^2(x)=v_0^2-(2/m)[V_{eff}(x)-V_{eff}(\pm \infty)]$ and $x_r$ is the reflection position given by $v_0^2=(2/m)[V_{eff}(x_r)-V_{eff}(\pm \infty)]$.\
 
In conclusion, by utilizing an effective particle approach, shown in good agreement with propagation simulation results, we have studied SS formation and LS dynamics at the interface between two dissimilar inhomogeneous media. The global trapping, reflection and transmission characteristics have been described in a direct analogy to Snell' laws for linear homogeneous media, while an analytic expression for the nonlinear analogue of the linear Goos-H\"{a}nchen shift has been obtained. It has been shown that different solitons can have qualitatively different dynamics in the same structure due to power (or width) dependent formation of effective potential barriers and wells. These effects could conceptually be considered as a power-dependent spatial filtering functionality.


\begin{thebibliography}{99}
\bibitem{JoViFa_07-Ru_03} J.D. Joannopoulos, P.R. Villeneuve and S. Fan, Nature \textbf{386}, 143149 (1997); P. Russel, Science \textbf{299}, 358 (2003). 
\bibitem{PhysRep_08} F. Lederer \textit{et al.}, Phys. Rep. \textbf{463}, 1 (2008).
\bibitem{DhLeSi_03-FlBaCo_05} D.N. Christodoulides, F. Lederer and Y. Silberberg, Nature \textbf{424}, 817 (2003); J.W. Fleicher \textit{et al.}, Opt. Express \textbf{13}, 1780 (2005).
\bibitem{BEC_1-2} B.P. Anderson and M.A. Kasevich, Science \textbf{282}, 1686 (1998); A. Trombettoni and A. Smerzi, Phys. Rev. Lett. \textbf{86}, 2353 (2001).
\bibitem{PeSuKi_04} D.E. Pelinovsky, A.A. Sukhorukov and Y.S. Kivshar, Phys. Rev. E \textbf{70}, 036618 (2004).
\bibitem{SiFiIl_08} Y. Sivan, G. Fibich and B. Ilan, Phys. Rev. E \textbf{77}, 045601 (2008).
\bibitem{KoHi_08} Y. Kominis and K. Hizanidis, Opt. Express \textbf{16}, 12124 (2008).


\bibitem{MaHuCh_06} K.G. Makris \textit{et al.}, Opt. Lett. \textbf{31}, 2774 (2006).
\bibitem{KaVyTo_06} Y.V. Kartashov, V.A. Vysloukh and L. Torner, Phys. Rev. Lett. \textbf{96}, 073901 (2006).
\bibitem{exp} S. Suntsov \textit{et al.}, Phys. Rev. Lett. \textbf{96}, 063901 (2006); C.R. \textit{et al.}, Phys. Rev. Lett. \textbf{97}, 083901 (2006); E. Smirnov \textit{et al.}, Opt. Lett. \textbf{31}, 2338 (2006).
\bibitem{KoPaHi_07} Y. Kominis, A. Papadopoulos and K. Hizanidis, Opt. Express \textbf{15}, 10041 (2007).
\bibitem{SuMaCh_0708}S. Suntsov \textit{et al.}, Opt. Express \textbf{15}, 4663 (2007); \textbf{16}, 10480 (2008).


\bibitem{AcMoNe_89} A.B. Aceves, J.V. Moloney and A.C. Newell, Phys. Rev. A \textbf{39}, 1809 (1989).
\bibitem{StSmRu_07} M. Stepic \textit{et al.}, Opt. Lett. \textbf{32}, 823 (2007).

\end{thebibliography}
\end{document}